# Quantum non-demolition measurement of a superconducting two-level system

A. Lupaşcu[1*], S. Saito[1,2], T. Picot[1], P. C. de Groot[1], C. J. P. M. Harmans[1] & J. E. Mooij[1]

[1]*Quantum Transport Group, Kavli Institute of NanoScience, Delft University of Technology, Lorentzweg 1, 2628 CJ Delft, The Netherlands*

[2]*NTT Basic Research Laboratories, NTT Corporation, 3-1, Morinosato-Wakamiya, Atsugi, Kanagawa, 243-0198, Japan*

[*]Present address: Laboratoire Kastler Brossel, Département de Physique de l'Ecole Normale Supérieure, 24 rue Lhomond, F-75231 Paris Cedex 05, France

**In quantum mechanics, the process of measurement is a subtle interplay between extraction of information and disturbance of the state of the quantum system. A quantum non-demolition (QND) measurement minimizes this disturbance by using a particular system – detector interaction which preserves the eigenstates of a suitable operator of the quantum system[1]. This leads to an ideal projective measurement. We present experiments in which we perform two consecutive measurements on a quantum two-level system, a superconducting flux qubit, by probing the hysteretic behaviour of a coupled nonlinear resonator. The large correlation between the results of the two measurements demonstrates the QND nature of the readout method. The fact that a QND measurement is possible for superconducting qubits strengthens the notion that these fabricated mesoscopic systems are to be regarded as fundamental quantum objects. Our results are also relevant for quantum information processing, where projective measurements are used for protocols like state preparation and error correction.**

Ever since the beginning of quantum mechanics, the interpretation of measurement has been a source of debate. In recent years, it has become possible to isolate, control, and measure single quantum objects and an accurate description of the inherent state disturbance associated with measurement has become even more important[1]. Quantum non-demolition (QND) detection is an important concept developed in this context. It was discovered during the search for optimal detection of gravitational waves[2] and has become an important paradigm of quantum physics. A QND method can be used for the measurement of a quantity which corresponds to an operator that is conserved during the free evolution of the quantum system. The interaction between the measurement apparatus and the measured system is designed in such a way that the eigenstates of this operator are not changed during the measurement. This strategy ensures that the apparatus will give an accurate result even when its response time is long compared to the characteristic dynamical time of the system. QND schemes have been demonstrated experimentally *e.g.* in quantum optics[3,4], atomic physics[5], and for a trapped single electron[6].

Mesoscopic superconducting systems with tunnel junctions exploit the nonlinearity of the Josephson effect and the large charging energy resulting from nanofabrication to create artificial two-level systems. Here the Hamiltonian can be readily controlled by applying control voltages and currents[7]. In recent years research on these systems was strongly motivated by their possible use as quantum bits. Quantum coherent behaviour of single qubits[8], entanglement of two qubits[9,10], and the coupling of a qubit to a harmonic oscillator mode[11-13] have been demonstrated. More recently, readout of single qubits with large fidelity was also realized[14-16]. However, a QND readout scheme is characterized not only by large fidelity, but also by the fact that the state of the qubit after the measurement is correlated with the measurement result. In the field of superconducting qubits, one possible route towards a QND type of readout is to employ the dispersive shift in a cavity-QED type of system[11]. In another approach, the qubit is coupled to a resonator by an interaction which is quadratic in the position-like coordinate of the resonator[17]. Exploiting the nonlinearity of such a resonator based on Josephson junctions, leads to fast and efficient detection of the qubit[15, 18, 19]. We use this latter approach for qubit state measurement. In the experiments we present here, large correlations of repeated measurements of a flux qubit are observed, providing a clear demonstration of a QND measurement for superconducting qubits.

Our superconducting qubit is a flux qubit[20,21]. It is formed of a micrometer sized superconducting aluminium loop, interrupted by Josephson junctions (see Fig. 1a). The Hamiltonian of the qubit $H_{qb} = h(\varepsilon(\Phi_{qb})\sigma_z + \Delta\sigma_x)/2$, where $h$ is Planck's constant and $\Delta$ is a fixed parameter related to the junction charging energies, is controlled by the externally applied magnetic flux in its loop $\Phi_{qb}$. The operators $\sigma_z$ and $\sigma_x$ have the usual Pauli matrix representation in the basis formed of two quantum states characterized by circulating persistent currents of equal amplitude $I_p$ and opposite sign. The parameter $\varepsilon = 2I_p(\Phi_{qb} - (n+1/2)\Phi_0)/h$, with $n$ the integer part of $\Phi_{qb}/\Phi_0$.

The two energy eigenstates of the qubit have the expectation value of the circulating current $\langle I_{circ} \rangle_{g,e} = \mp \varepsilon/\sqrt{\varepsilon^2 + \Delta^2} I_p$ for the ground ($g$) and excited ($e$) state, respectively. The measurement is realized with a DC-SQUID magnetometer, coupled to the qubit by a mutual magnetic inductance $M$, which senses the difference in current between the states $g$ and $e$. The SQUID is characterized by a nonlinear inductance $L_J$, with a value which depends on the qubit state. The difference between the values of the inductance $L_J(g)$ and $L_J(e)$, corresponding to the qubit being in the $g$ or $e$ states respectively, is ~4 % in our experiments. This difference is detected by including the Josephson inductance $L_J$ in a resonant circuit, formed by the addition of a microfabricated capacitor $C$ (see Fig. 1b). An applied microwave signal of power $P_{probe}$ and frequency $F$ close to the resonance frequency $F_{res}$ is reflected with a phase which depends on the inductance $L_J$ and thus on the qubit state.

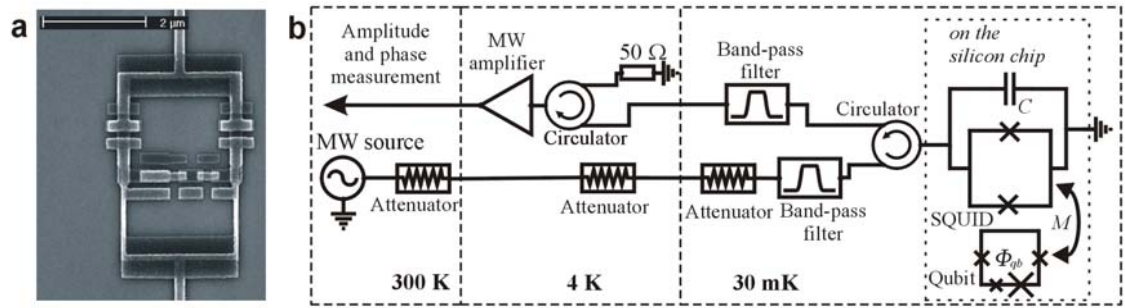

**Figure 1**. The flux qubit and the measurement system. **a**, The qubit and the DC-SQUID are fabricated on an oxidized silicon substrate using electron-beam lithography. The Josephson tunnel junctions are formed by the overlap of two aluminium layers with thickness of 25 nm and 50 nm respectively, separated by a thin aluminium oxide layer. The qubit is the lower loop. It contains two junctions with equal area characterized by a Josephson energy $E_J$=250 GHz and a charging energy $E_C$=6.7 GHz, and a third junction smaller by a factor $s$=0.73. The experimentally determined qubit parameters are $\Delta$=4.6 GHz and $I_p$=340 nA, in good agreement with the mentioned values of $E_J$, $E_c$, and $s$. A fourth auxiliary large junction ensures symmetric coupling of the qubit and SQUID. The DC-SQUID, formed by the combined loop, has two junctions with critical current $I_{c0}$=2 μA. The mutual inductance $M$=14 pH represents the sum of a geometric inductance and of the kinetic inductance of the narrow lines shared by the qubit and SQUID loops[21]. **b**, The resonator used for qubit readout is formed by the SQUID inductance $L_J$=219 pH, the microfabricated capacitor $C$=37 pF, and a stray series inductance $L_S$=73 pH. A probe microwave is sent to the resonator through a circulator and the reflected wave is amplified using a low-noise cryogenic amplifier with a noise temperature $T_n$~3 K. A second circulator, placed at 4 K, further suppresses noise reaching the qubit –

resonator circuit. The quadratures of the reflected wave are measured, which allows the detection of both amplitude and phase.

The interaction between qubit and resonator is described by the Hamiltonian $H_{int} = k\sigma_z(a+a^*)^2$, where $\sigma_z$ is a qubit operator, $a^*$ ($a$) is the creation (annihilation) operator for the resonator and $k$ is the qubit-resonator coupling constant. The most restrictive criterion for QND measurement requires that $[H_{qb}, H_{int}] = 0$ [1]. This condition ensures that the interaction of the qubit with the detector does not result in a disturbance of the probability for the qubit to be in either the $g$ or $e$ state. For our system, this commutation relation is not strictly satisfied. However, the large difference between the frequencies of the oscillator and the qubit, as well as the setting $\varepsilon \gg \Delta$ used in the measurement, ensures that the probability of transitions between $e$ and $g$ is very small, allowing non-demolition detection.

Strong driving of the nonlinear resonator results in a bistable behaviour with hysteresis, which can be used for high qubit readout efficiency. The resonator can be described by the classical Duffing model[22]. For a certain range of driving conditions (see Fig. 2a) the resonator can reside in any of the two possible states, denoted by $l$ and $h$, with different oscillation amplitude and phase. These states can be distinguished by the phase of the reflected microwave used to probe the resonator. The measurement shown in Fig. 2b illustrates this behaviour. Transitions between the states $l$ and $h$ are observed when the resonator driving power $P_{probe}$ approaches the bifurcation thresholds $P_{probe}^{high}$ and $P_{probe}^{low}$. These transitions depend exponentially on the difference between the driving power $P_{probe}$ and the bifurcation thresholds. The strong dependence of the rate of the transition $l \rightarrow h$ on the upper bifurcation threshold $P_{probe}^{high}$ is used for qubit readout, as proposed by Siddiqi et al.[23] and demonstrated in experiments[15, 18, 19]. $P_{probe}^{high}$ depends on the inductance $L_J$, and thus on the qubit state (see Fig. 2c). To further improve the qubit readout, we modulate the amplitude of the driving current as indicated in Fig. 2d. The level and duration of the first part of the pulse (*switching*) is chosen such that the probability for the $l \rightarrow h$ transition is nearly 0 % (100 %) when the qubit is in the *e* (*g*) state. The second part of the pulse (*latching*) has a reduced amplitude such that both transitions $l \rightarrow h$ and $h \rightarrow l$ have negligibly small probabilities. The latching time is chosen just long enough that the discrimination of the states *h* and *l* against electrical noise is possible.

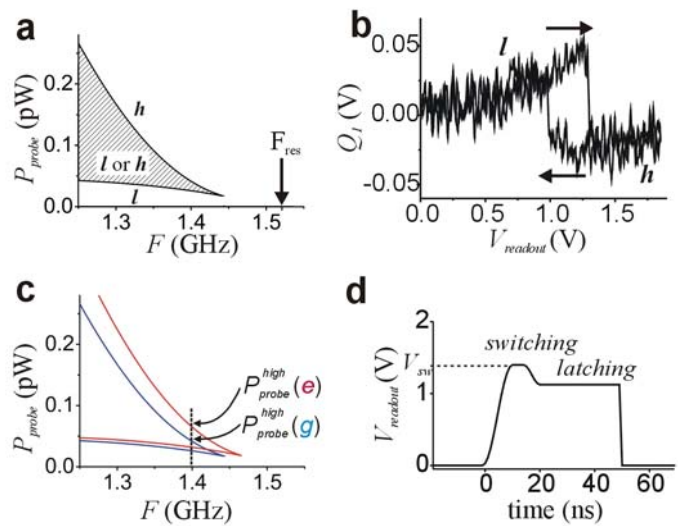

**Figure 2**. Bistability of the driven nonlinear resonator and qubit state detection. **a,** Stability diagram of the driven resonator. In the hatched region two oscillation states are possible. Below the hatched region, only the state *l*, with low oscillation amplitude is present. Above, only the state *h*, with high amplitude, is possible. **b,** One amplified quadrature, $Q_1$, of the reflected signal versus $V_{readout}$. $V_{readout}$ is a control voltage that sets the amplitude of the probing power; $P_{probe}$ [pW] ~ 0.04 ($V_{readout}$[V])$^2$. $V_{readout}$ is ramped linearly first from 0 V to 1.85 V in 5 μs and then from 1.85 V to 0V in 5 μs. **c,** Limits of the bistability region for the cases of qubit in state *g* (blue) and *e* (red). At the frequency $F$ =1.4 GHz, used in the measurements, the bifurcation thresholds $P_{probe}^{high}(g)$ and $P_{probe}^{high}(e)$

differ significantly. **d,** Modulation of the amplitude of the resonator driving signal for a typical readout pulse. We denote the amplitude of the switching plateau by $V_{sw}$.

A typical sequence for the manipulation and readout of the qubit is shown in Fig. 3a. We prepare the qubit in the ground state with large fidelity by waiting for a time much longer than the qubit relaxation time. This is possible because the energy level splitting $h\nu_{qb}$, with $\nu_{qb}$ = 14.2 GHz, is much larger than the effective temperature $T$ < 100 mK. Next we excite the qubit coherently by applying an AC magnetic flux $\delta\Phi_{qb}$ at frequency $\nu_{qb}$ for a time $\Delta t$. This pulse induces Rabi oscillations between the states $g$ and $e$. Finally, qubit measurement is performed by switching on the driving of the resonator and detecting the state of the latter to be $l$ or $h$. This sequence is repeated at least $10^4$ times and the probability for the resonator to be in state $h$, $P(h)$, is determined. In Fig. 3b we plot this probability as a function of the amplitude of the switching plateau of the readout pulse $V_{sw}$, while keeping the amplitude of the latching plateau constant. This measurement is done for two qubit states, $g$ and $\tilde{e}$, where the state $\tilde{e}$ corresponds to the optimal preparation of the excited state by a Rabi pulse. The maximum difference between $P(h)$ for the $g$ and $\tilde{e}$ states is 85.4 %, as indicated in Fig. 3b. Ideally, when the qubit is in state $g$ ($e$), we detect the oscillator in the state $h$ ($l$). Imperfection of the detection is described by the parameters $\alpha$ and $\beta$ (see Fig. 3d). Our measurements indicate that $\alpha$ <0.2 % and $\beta$ <15 % . An accurate determination of $\alpha$ and $\beta$ is hampered by the failure to prepare the state $\tilde{e}$ very close to $e$, due to decoherence during the Rabi oscillations (see Fig. 3c). With the assumption that the ground state is prepared with large fidelity and that the observed decay of the Rabi oscillation is exponential (see Fig. 3c) we can infer the values $\alpha$ = (0.2 ± 0.1) % and $\beta$ = (9.5 ± 0.8) %.

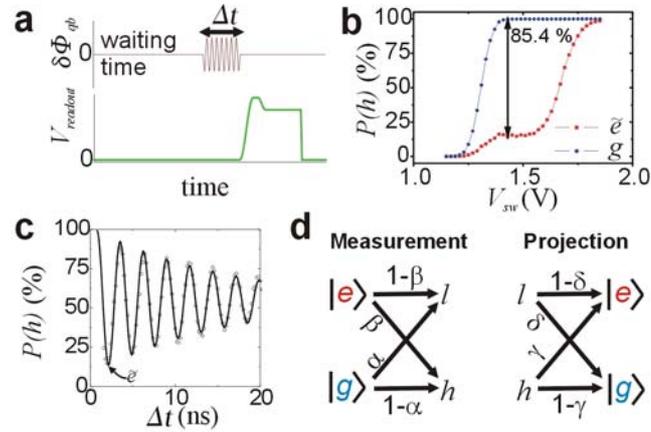

**Figure 3**. Measurement of qubit state. **a**, Schematic representation of the qubit excitation pulse (top) and readout pulse (bottom) sequence. **b**, Probability to detect the oscillator in the state *h* for the qubit states *g* (blue dots) and $\tilde{e}$ (red squares) for 16384 repetitions of the measurement; the solid lines are guides to the eye. For $V_{sw}$=1.43 V, the difference between the two probabilities is maximum, equal to 85.4 %. The readout pulse latching plateau is 150 ns long, which insures that the states *l* and *h* are discriminated with an error smaller than 0.1 %. **c**, Measurement of Rabi oscillations (circles) and a fit with an exponentially damped sinusoidal shape (solid line). The label indicates the Rabi pulse time which produces the qubit state $\tilde{e}$. **d,** Schematic representation of the parameters characterizing the measurement errors.

The QND character of the readout is analyzed by preparing the qubit state and then performing two consecutive measurements, as illustrated in Fig. 4a. To limit qubit relaxation during the first measurement, the first latching plateau has a duration of only 30 ns. With this duration the error in the discrimination of the resonator states *l* and *h* becomes ~ 0.3 %, increasing the values of the measurement error parameters to $\alpha$ = (0.5 ± 0.1) % and $\beta$ = (9.8 ± 0.8) %. We define $P(m_2|m_1)$ as the conditional probability to obtain the result $m_2$ in the second measurement if the result of the first measurement was $m_1$. The measured $P(l|l)$ and $P(h|h)$ are very high if the two measurements follow immediately after each other (see Fig. 4b). The correlation $P(l|l)$ decreases with the delay time between measurements. The observed decay is exponential with a characteristic time equal to the qubit relaxation time, measured independently. These results are consistent with the following picture: if the first measurement yields the result $l$, the qubit is projected to state $e$, and subsequently decays to the $g$ state. This decay results in a reduced probability to obtain $l$ in the second measurement. $P(h|h)$ is nearly constant because the first measurement results in the qubit state $g$ and no energy relaxation will occur.

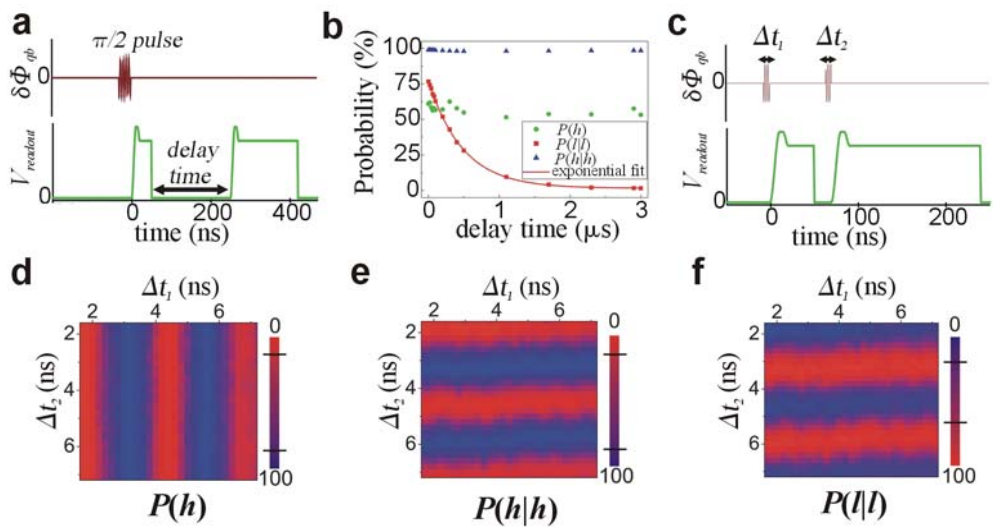

**Figure 4**. Correlation measurements of the qubit. **a,** Schematic representation of the qubit control (top) and readout (bottom) pulses used to determine the dependence of conditional probability on the delay time between two measurements. **b,** Probability *P(h)* for the first measurement (green dots), conditional probability *P(h|h)* (blue triangles), and conditional probability *P(l|l)* (red squares) with an exponential fit (red solid line). **c,** Schematic representation of the qubit control (top) and readout (bottom) pulses used to measure *P(h)*, *P(h|h)*, and *P(l|l)* as a function of the durations $\Delta t_1$ and $\Delta t_2$ of microwave pulses

inducing Rabi oscillations before the first and the second measurement. **d,e,f,** Measurements of *P(h)*, *P(h|h)*, and *P(l|l)*, respectively, with the procedure shown in **c**. The black lines on the scale bar indicate the minimum and maximum value for each dataset.

In order to clarify whether the readout is projective, it is required to measure the correlations *P(h|h)* and *P(l|l)* for arbitrary initial qubit states as well as to verify that the measured correlations are a result of qubit state projection and not an artefact of our measurement procedure. We therefore employ the control and readout sequence shown in Fig. 4c. The probability *P(h)* for the first measurement varies sinusoidally with $\Delta t_1$ as a result of induced qubit Rabi oscillations and is completely independent of $\Delta t_2$ (see Fig. 4d), due to the large waiting time between repetitions of the sequence. The conditional probabilities *P(h|h)* and *P(l|l)* are independent of $\Delta t_1$ (see Figs. 4e,f) although the qubit may have many possible initial states, as shown by the result of the first measurement (see Fig. 4d). This illustrates the strong projective character of the measurement. We note that two previous experiments, on a flux qubit[15] and on a phase qubit[24], illustrate the case of weak partially projective measurements where the correlations do depend on the initial qubit state. The dependence of *P(h|h)* and *P(l|l)* on the length $\Delta t_2$ of the second Rabi pulse confirms that the measured correlations are indeed due to the qubit and not to spurious dynamics of the detector.

In Fig. 3d, we represent the parameters $\gamma$ and $\delta$, which describe the non-ideal character of state projection. We define the QND fidelity to be the probability that the qubit initial state is preserved after measurement, irrespective of the measurement outcome, when the initial state is either *g* or *e* with equal probability. It is given by $F_{QND} = (1 + (1 - \alpha - \beta)(1 - \gamma - \delta))/2$, which can be expressed in terms of the measured conditional probabilities as $F_{QND} = (P(l|l) + P(h|h))/2$, yielding $F_{QND}$ = 88 %. By using the values $\alpha$ and $\beta$ determined as explained above, and the measured values of

$P(h|h)$ and $P(l|l)$, we determine $\gamma = (0.6 \pm 0.1)$ % and $\delta = (14.6 \pm 0.9)$ %. The large value of $\delta$ is due to relaxation of the qubit during the latching plateau of the first readout pulse. In the absence of resonator driving the qubit relaxation time is $T_1$ =470 ns (see Fig. 4b), decreasing to 260 ns when the oscillator is driven. By taking into account this relaxation, we can calculate the projection error parameters $\gamma$ and $\delta$ that refer to the qubit state immediately after the switching part of the readout pulse yielding $\gamma_{corrected}$ = $(1.7 \pm 0.3)$ % and $\delta_{corrected}$ = $(4.2 \pm 1)$ %. With this correction, the QND fidelity would be $F_{QND,corrected}$ = 92 %.

Our experiments demonstrate QND detection for a superconducting flux qubit, with a measured QND fidelity of 88 %. The measurement scheme results in highly effective state projection, as prescribed in quantum mechanics textbooks but difficult to achieve in practice. As a consequence this readout method is relevant for detection-induced state preparation as well as error correction in quantum information processing.

**Acknowledgements** We thank M. Devoret, P. Bertet, T. Meunier, G. Nogues, and J.M. Raimond for discussions, and R. N. Schouten for his contribution to the measurement system. Financial support was provided by the Dutch Foundation for Fundamental Research on Matter (FOM), the EU projects SQUBIT2 and EuroSQIP, and the Dutch National Initiative on Nano Science and Technology NanoNed.

**Competing Interests statement** The authors declare that they have no competing financial interests.

**Correspondence** and requests for materials should be addressed to A.L. (Adrian.Lupascu@lkb.ens.fr).